\documentclass[prd,11pt,aps,superscriptaddress,floatfix,showkeys]{revtex4}
\usepackage{graphicx,amsmath,bbm,bm,array}
\usepackage[linktocpage=true,colorlinks=true,linkcolor=blue,citecolor=blue]{hyperref}

\begin{document}
\title{Electrical conductivity of a hot and dense QGP medium in a magnetic field}
\author{Lata Thakur\footnote{Present address : Asia Pacific Center for Theoretical Physics, Pohang 37673, Korea}}
\email{lata.thakur@apctp.org} 
\affiliation{School of Physical Sciences, National Institute of Science Education and Research, HBNI, Jatni 752050, India}
\author{P. K. Srivastava} 
\email{prasu111@gmail.com}
\affiliation{ Department of Physics, Indian Institute of Technology Ropar, Ropar 140001, India}

\begin{abstract}
	We compute the electrical conductivity ($ \sigma_{el} $) in the presence of constant and homogeneous external electromagnetic field for the static quark-gluon plasma (QGP) medium, which is among the important transport coefficients of QGP. We present the derivation of the electrical conductivity by solving the relativistic Boltzmann kinetic equation in the relaxation time approximation in the presence of magnetic field ($ B $). We investigate the dependence of electrical conductivity on the temperature and finite chemical potential in magnetic field. We find that electrical conductivity decreases with the increase in the presence of magnetic field. We observe that $ \sigma_{el} $ at a nonzero $ B $ remains within the range of the lattice and model estimate at $ B\neq 0 $. 
	 Further, we extend our calculation at finite chemical potential.
\end{abstract}


\maketitle 
\section{Introduction}
\noindent
The ongoing experiments at the Relativistic Heavy Ion Collider (RHIC) at BNL and at the Large Hadron Collider (LHC) at CERN are aimed to produce a new state of matter at high temperature and/or baryonic density which is known as quark-gluon plasma. The study of various properties of this hot and dense QCD medium became a topic of great interest. Transport coefficients such as the shear and bulk viscosity and diffusion constants 
are of fundamental importance 
in the dynamics of the formation and evolution of QCD matter.
Electrical conductivity is important
among the various transport coefficients, which has been calculated by several research groups for the QCD matter \cite{Arnold:2000dr,Arnold:2003zc,Gupta:2004,Aarts:2007wj,Buividovich:2010tn,Ding:2010ga,Burnier:2012ts,Brandt:2012jc,Amato:2013naa,Aarts:2014nba,Cassing:2013iz,Steinert:2013fza,Hirono:2012rt,Greif:2014oia,Puglisi:2014pda,Puglisi:2014sha,Finazzo:2013efa,Greif:2016skc,Mitra:2016zdw,Srivastava:2015via,Thakur:2017hfc,Marty:2013ita,Fernandez-Fraile2006,Kadam:2017iaz}.
 
Recently, it was suggested that an extremly strong magnetic field is generated in noncentral heavy ion collisions, which is of 
great phenomenological significance. The
magnitude of the magnetic field is estimated to be of the
order of or 
beyond the
intrinsic QCD scale
 $\Lambda_{\text {QCD}} $ ($ \Lambda^{2}_{\text {QCD}} \leq eB $) and is reaching several $ m_{\pi}^{2} $ within proper time $ 1-2 $ fm/c~\cite{Tuchin:2013ie,Kharzeev:2007jp}.
Therefore, it is becoming important to understand the various aspects of quark-gluon plasma (QGP) in the presence of magnetic field.
Several phenomena like magnetic catalysis \cite{Shovkovy:2012zn}, the chiral magnetic effect \cite{Fukushima:2008xe,Kharzeev:2007jp,Kharzeev:2009fn,Kharzeev:2010gr,Huang:2015fqj,Warringa:2012bq}, the chiral magnetic wave~\cite{Kharzeev:2010gd}, and anomalous charge separation~\cite{Huang:2015fqj}
 occur in the presence of magnetic field. The effect of magnetic field has been studied within the framework of relativistic magnetohydrodynamics \cite{Inghirami:2016iru,Roy:2015kma}. Its effects has also been investigated for the flow anisotropy \cite{Das:2016cwd,Mohapatra:2011ku,Feng:2017giy}, the heavy-quark diffusion constant~\cite{Fukushima:2015wck}, the heavy quark potential~\cite{Bonati:2015dka,Bonati:2016kxj,Singh:2017nfa,Hasan:2017fmf}, shear viscosity\cite{Nam:2013fpa,Mohanty:2018eja} and the bulk viscosity~\cite{Hattori:2017qih,Kurian:2018dbn} of the QGP medium.
 
In the present work, we investigate the effect of constant and homogeneous magnetic field on the electrical conductivity of the static QGP medium. The electrical conductivity, $\sigma_{el} $, of the medium is intimately related to the evolution of the
electromagnetic field in a conducting plasma \cite{Das:2017qfi,Tuchin:2010gx}. The magnetic field produced in heavy ion collision can be sustained in the QGP medium with finite electrical conductivity as discussed in Refs.~\cite{Mohapatra:2011ku,Tuchin:2010gx}.  It is argued that initially the magnetic field
decreases very fast, but due to the matter effects at later times the decrease of magnetic field is very slow. In addition, the electrical conductivity in the presence of magnetic field for the QCD matter has been studied by different approaches like perturbative QCD~\cite{Hattori:2016lqx}, the kinetic theory approach and kubo formula \cite{Feng:2017tsh}, the effective fugacity quasiparticle model~\cite{Kurian:2019fty}, the quasiparticle model~\cite{Rath:2019vvi}. Electrical conductivity 
has also been studied for the hot and dense hadronic matter in a
magnetic field using the hadron resonance gas model~\cite{Das:2019wjg}.
Here, we study the electrical conductivity of the hot QGP medium in the presence of magnetic field by using a quasiparticle model~\cite{Goloviznin:1992ws,Peshier:2002ww,Bannur:2006ww,Bannur:2006js,Peshier:1995ty,Peshier:1999ww,Srivastava:2010xa,Srivastava:2012bv} within the framework of relaxation time approximation. This model provides a reasonable transport and thermodynamical behavior of the QGP phase.  

This article is organized as follows. In the next section, we calculate
the electrical conductivity in the presence of magnetic field by using the Boltzmann equation in the relaxation time approximation (RTA). In Sec.
\ref{secIII}, we discuss the quasiparticle model. Further, in Sec. \ref{secIV}, we
present our results regarding the electrical conductivity and compare
with the lattice as well as other phenomenological calculations. Finally, we extend our results at finite chemical potential and give the conclusion drawn from our work.

\section{Electrical Conductivity}
\label{secII} 
The relativistic Boltzmann transport (RBT) equations for relativistic particle in the presence of an external electromagnetic field in the RTA is given by~\cite{Hosoya:1983xm,Yagi}
\begin{equation} 
\left(k^{\mu}\partial_{\mu}^{x} + F^{\mu}{\partial_{\mu}^{k}}\right) f(x,k,t) = -\frac{k^{\mu}u_{\mu}}{\tau}( f-f_0),
\label{RBTeq}
\end{equation}
where $u^\mu=(1,0)$ is the fluid four velocity and $ F^{\mu} $ is the external force defined as
  $F^{\mu}=(k^0v {\bf .F},k^0 {\bf F})  $.\\
In the classical electrodynamics, for the Lorentz force, we have $ {\bf F}=q ({\bf E}+v\times {\bf B}) $, where $ q $ is the charge of particles, with $ E $ and $ B $ the electric and magnetic fields. By using the relation $F^{0i}=E^{i}$ and $ 2F_{ij}=\epsilon_{ijk}B^k $, where $ F^{\mu\nu} $ is the (antisymmetric) electromagnetic field tensor, we find that
\begin{equation}
F^{\mu}=-qF^{\mu\nu}k_{\nu}
\end{equation}
and also
\begin{equation}
k_{\mu}F^{\mu}=0,~~~~~~\textit{and}~~~~~{\partial_{\mu}^{k}}F^{\mu}= 0.
\label{F0}
\end{equation}
It is instructive to consider the relativistic Boltzmann equation (\ref{RBTeq}) in three-notation. Then, we get
\begin{equation}
\left(k^{0}\partial_{0}^{x}+ k^{i}\partial_{i}^{x} + F^{0}{\partial_{0}^{k}} +F^{i}{\partial_{i}^{k}}\right) f(k) = -\frac{k^{0}}{\tau}( f-f_0),
\end{equation}
\begin{equation}
\left(\dfrac{\partial f}{\partial t}+{\bf v}.\dfrac{\partial f}{\partial {\bf r}} + \frac{{\bf F.k}}{k^0}\frac{\partial f}{\partial { k^{0}}} +{\bf F}.\frac{\partial f}{\partial {\bf k}}\right) = -\frac{( f-f_0)}{\tau},
\end{equation}
Here we are considering that the spatially uniform field is applied to a steady-state system such that there are no space-time gradients~\cite{Feng:2017tsh}. For a spatial homogeneous distribution function, $ \frac{\partial f}{\partial {\bf r}}\approx 0 $, and for the steady-state condition,
$ \frac{\partial f}{\partial t}=0 $ . 
Therefore, 
we get
\begin{equation}
\left(\frac{{\bf F.k}}{k^0}\frac{\partial}{\partial { k^{0}}} +{\bf F}.\frac{\partial}{\partial {\bf k}}\right) f(k) = -\frac{( f-f_0)}{\tau},
\end{equation}
The chain rule of differentiation implies
\begin{equation}
\frac{\partial k^{0}}{\partial {\bf k}}\frac{\partial}{\partial { k^{0}}}+\frac{\partial}{\partial {\bf k}}\rightarrow \frac{\partial}{\partial {\bf k}},
\end{equation}
and we have
\begin{equation}
{\bf F}.\frac{\partial}{\partial {\bf k}} f(k) = -\frac{( f-f_0)}{\tau}
\end{equation}
\begin{equation}
q ({\bf E}+v\times {\bf B}).\frac{\partial}{\partial {\bf k}} f(k) = -\frac{( f-f_0)}{\tau}
\label{RBT8}
\end{equation}
To further simplify the above RBT equation we consider $ {\bf E}=E\hat{x} $ and $ {\bf B}=B\hat{z} $. Then we have
\begin{equation}
f-qB\tau \left(v_{x}\frac{\partial f}{\partial  k_{y}}- v_{y}\frac{\partial f}{\partial  k_{x}
}    \right)=f_{0}-qE\tau \frac{\partial f_{0}}{\partial  k_{x}}
\label{RBT9}
\end{equation}
To solve Eq.(\ref{RBT9}), we take the following ansatz of the distribution function $f(k)$ ~\cite{Feng:2017tsh}
\begin{equation}
f(k)=f_0-\tau q{\bf E}.\frac{\partial f_0(k)}{\partial {\bf k}}-{\bf \Xi}.\frac{\partial f_0(k)}{\partial {\bf k}}.
\label{ansatz}
\end{equation}
Here $  f_{ 0} $ is the equilibrium
distribution and is given by the Fermi Dirac distribution as
\begin{equation}
f_{0}({\bf k})=\frac{1}{e^{(\sqrt{{\bf k}^{2}+m^{2}}\pm \mu)/T}+1}, 
\end{equation}
which is a space- and time- independent solution to the Boltzmann equation. 
Using the ansatz given in Eq. (\ref{ansatz}), we can simplify Eq.~(\ref{RBT9}), 
\begin{equation}
\tau qB qE\frac{v_y}{\epsilon}-qB\bigg(v_x\Xi_y-v_y\Xi_x\bigg)\frac{1}{\epsilon}+\frac{1}{\tau}\bigg(\Xi_x\frac{k_x}{\epsilon}+\Xi_y\frac{k_y}{\epsilon}
+\Xi_z\frac{k_z}{\epsilon}\bigg)=0,
\label{equ9}
\end{equation}
where $ \epsilon=\sqrt{{\bf k^{2}}+m^2} $. The above equation should be satisfied for any value of the velocity, hence, one can get $\Xi_z=0$. After comparing the coefficients of $v_x$ and $v_y$ on both sides 
of Eq.~(\ref{equ9}), one can get 
\begin{equation}
\omega_c\Xi_x+\frac{\Xi_y}{\tau}+\omega_c(qE\tau)=0,
\label{sol1}
\end{equation}
and
\begin{equation}
\frac{1}{\tau}\Xi_x-\omega_c\Xi_y=0,
\label{sol2}
\end{equation}
where $\omega_c=qB/\epsilon$ is the cyclotron frequency. Solving Eqs. (\ref{sol1}) and (\ref{sol2}) for $\Xi_x$ and $\Xi_y$, one obtains 
\begin{equation}
\Xi_x=-\frac{\omega_c^2\tau^3}{(1+\omega_c^2\tau^2)}qE, ~~~\Xi_y=-\frac{\omega_c\tau^2}{1+\omega_c^2\tau^2}qE.
\label{equ12}
\end{equation}
Therefore, by using Eq.~(\ref{equ12}), the ansatz for the distribution function $f(k)$ given in Eq.~(\ref{ansatz}) can be simplified as
\begin{eqnarray}
f(k) =f_0-\frac{qEv_x\tau}{1+\omega_c^2\tau^2}\left(\frac{\partial f_0}{\partial \epsilon}\right)
+\frac{qEv_y\omega_c\tau^2}{1+\omega_c^2\tau^2
}\left(\frac{\partial f_0}{\partial\epsilon}\right)
\label{equ13}
\end{eqnarray}
and we can obtain $ \delta f $ as
\begin{equation}
\delta f=f-f_0=-\frac{qEv_x\tau}{1+\omega_c^2\tau^2}\left(\frac{\partial f_0}{\partial \epsilon}\right)
+\frac{qEv_y\omega_c\tau^2}{1+\omega_c^2\tau^2
}\left(\frac{\partial f_0}{\partial\epsilon}\right)
\label{delf}
\end{equation}
The electrical conductivity ($ \sigma_{el}$) represents the response of
the system to an applied electric field ({\bf E}). From Ohm's
law electric current ({\bf J}) can be written in terms of $ \sigma_{el}$ as
\begin{equation}
\bf J= \sigma_{el} \bf E.
\end{equation}
The electric four current ($ J^{\mu} $) can be written as
\begin{eqnarray}
J^{\mu}=  g\int \frac{d^{3}k}{(2 \pi)^3\epsilon} k^{\mu}\lbrace qf(k)- {\bar q}{\bar f(k)}\rbrace,
\label{current}
\end{eqnarray}
where $ q $ (${\bar q} $) is the charge for quarks (antiquarks). 
Equation (\ref{current}) at the zero chemical potential ($ \mu=0 $) reduces to 
\begin{equation}
J^{\mu}= 2q g \int \frac{d^{3}k}{(2 \pi)^3\epsilon} k^{\mu}f(k).
\end{equation}
In the presence of some external electromagnetic field, $ J^{\mu}=J_{0}^{\mu} + \Delta J^{\mu}$, where
\begin{equation}
\Delta J^{\mu}= 2q g \int \frac{d^{3}k}{(2 \pi)^3\epsilon} k^{\mu} \delta f.
\label{delj}
\end{equation}
Using $ \delta f $ from Eq.~(\ref{delf}), generalizing to a system of different charged particles and considering the definition of electrical conductivity, 
one obtains 
\begin{equation}
\sigma_{\rm{el}} = \frac{1}{3\pi^2 T} \sum_f 
g_{{}_f} q_{{}_f}^{2} \int dk \frac{k^4}{\epsilon_{f}^2}  
\frac{\tau_{{}_f}}{(1+\omega_c^2\tau_{{}_f}^2)} f_{{}_f}^{0}(1-f_{{}_f}^{0}).
\label{siginB}
\end{equation}
Electrical conductivity at finite chemical potential ($ \mu\neq0 $)
\begin{equation}
\sigma_{\rm{el}} = \frac{1}{6\pi^2 T} \sum_f 
g_{{}_f} q_{{}_f}^{2} \int dk \frac{ k^4}{\epsilon_{f}^2}  
\frac{\tau_{{}_f}}{(1+\omega_c^2\tau_{{}_f}^2)} \left[ f_{{}_f}^{0}(1-f_{{}_f}^{0})-\bar{f_{{}_f}}^{0}(1-\bar{f_{{}_f}}^{0})\right].
\label{sigatmu}
\end{equation}
In the relaxation time approximation, the system is not very far from equilibrium. Hence we assume that the quark distribution function is always close to equilibrium and
introduce very small deviations from the equilibrium, which allows the
linearization of the RBT equation. 
This shows
that the magnetic field, $ {\bf B} $ cannot be very strong. Therefore, we are not considering the Landau quantization of the charged particle in a magnetic field. 

\section{Quasiparticle Model}
\label{secIII}
The quasiparticle model 
 is a phenomenological model which can be applied to study the thermal properties of QGP at physically relevant, low temperatures near the phase
 transition temperature, $ T_{c} $, where one cannot make use of perturbative QCD directly. The nonperturbative effects become important at low temperature, near the phase transition point, where the first principle lattice calculations become reliable. However, one needs an effective description of QGP near $ T_{ c} $ for phenomenological models. Since even at relatively low temperature the gas of quasiparticle still remains weakly interacting, one can treat this gas in a perturbative way down to critical temperature.
 In this model, the system of
 interacting massless partons (quarks and gluons) can be effectively
 described as an ideal gas of massive noninteracting
 quasiparticles. The mass of these quasiparticles, $ m_{th}(T) $, depends on the temperature and arises due to the interactions of quarks and gluons with the surrounding medium. Such a functional dependence of thermal mass turns out to reproduce the lattice data quite well at finite temperature. This model was first proposed by Goloviznin and Satz \cite{Goloviznin:1992ws} and then by Peshier {\it et al.} \cite{Peshier:1995ty,Peshier:2002ww} to explain the equation of states
 of QGP obtained from lattice gauge simulation of QCD at
 finite temperature. Simultaneously,
 different authors in Refs.~\cite{Plumari:2011mk,Bluhm:2004xn,Bluhm:2007nu,Bluhm:2007cp} discussed the high-temperature lattice data
 by using a suitable quasiparticle description for QGP in
 which the constituents of QGP medium acquire a $ T $- and/or
 $ \mu $- dependent mass. These results suggest that the high-temperature QGP phase is suitably described by a thermodynamically consistent quasiparticle model. This model has been found to work well above and around
 the critical temperature $ T_{ c} $.
All the quarks have both the bare mass, $m_{i0}$, and the thermal mass, $m_{\text {th}}$, and hence the expression for the effective mass of quarks and antiquarks is 
~\cite{Bannur:2006ww,Srivastava:2010xa,Srivastava:2012bv}
\begin{equation}
m_{i}^{2}=m_{0i}^{2}+\sqrt{2}m_{0i}m_{th,i}+m_{th,i}^{2},
\label{mth1}
\end{equation}
where $m_{\text {th}}$, which arises due to the interaction of quarks (antiquarks) with the constituents of the medium, can be written as~\cite{Peshier:2002ww,Peshier:1999ww,Braaten:1991gm}
\begin{equation}
m_{th,i}^{2}=\frac{g^{2}T^{2}}{6}\left(1+\frac{\mu_{i}^{2}}{\pi^{2}T^{2}}\right),
\label{mth}
\end{equation}
where $ g^{2} =4\pi\alpha_{s}$, and the strong coupling constant, $\alpha_{s} $ for one loop in the presence of magnetic field is given by~\cite{Ayala:2018wux,Bandyopadhyay:2017cle}
\begin{equation}
\alpha_{s}(\Lambda^{2},|eB|)=\frac{\alpha_{s}(\Lambda^{2})}{1+b_{1}\alpha_{s}(\Lambda^{2})\ln\left(\frac{\Lambda^{2}}{\Lambda^{2}+|eB|}\right)}
\end{equation}
and the one-loop running coupling at $ B=0 $ is
\begin{equation}
\alpha_{s}(\Lambda^{2})=\frac{1}{b_{1}\ln\left(\frac{\Lambda^{2}}{\Lambda_{\overline{\rm MS}}^{2}}\right)}
\label{alpha}
\end{equation}
where $ b_{1}=\frac{(11N_{c}-2 N_{f})}{12 \pi} $ and $\Lambda_{\overline{\rm MS}}=176 $ MeV for $ N_f=3 $. Here for quarks, $ \Lambda=\Lambda_q=2\pi\sqrt{T^2+\mu^2/\pi^2} $ and for gluons, $\Lambda= \Lambda_g=2\pi T $. The thermal mass depends on the QCD coupling constant and here we found that $ \alpha_{s} =400$ MeV at $ T=160 $ MeV. 
Thus, the Eqs~(\ref{mth}) and ~(\ref{alpha}) are valid at this temperature and we can extrapolate the quasiparticle model (QPM) results at $ T=160 $ MeV.

In Eq.~(\ref{siginB}) and~(\ref{sigatmu}), $ \tau_{f} $ is the relaxation time for quarks, antiquarks, and gluons that can be calculated by using the following expressions as given in Ref.~\cite{Hosoya:1983xm} for the massless case
\begin{equation}
\tau_{q(\bar{q})}=\frac{1}{5.1 T\alpha_{s}^{2} \log \left(\frac{1}{\alpha_{s}}\right)\left(1+0.12 (2 N_{f}+1)\right)}
\end{equation}
\begin{equation}
\tau_{g}=\frac{1}{22.5 T \alpha_{s}^{2} \log \left(\frac{1}{\alpha_{s}}\right)\left(1+0.06 N_{f}\right)}.
\label{tau}.
\end{equation}
For simplicity, the relaxation time has been used for the massless case.
It is clear from the Ref.~\cite{Berrehrah:2013mua}, that the effect of the massive quark is small in the estimation of the scattering cross-sections, which results in a negligible effect on the relaxation time. Therefore our results remain almost same for the massive particles case as well.‘

In the QPM, partons are treated as particles having rest as well as thermal mass [Eq.~ (\ref{mth1})].
Thus, the distribution function of the QPM contains both the rest as well as thermal mass. 
Here we take the rest mass of the up ($ u $),down ($ d $) and strange quarks as $  m _{0u(d)} =8$~MeV, and $  m _{0s} =80$~MeV,~\cite{Srivastava:2010xa}.

\section{Results and Discussions}\label{secIV}
\begin{figure}
\includegraphics[scale=0.45]{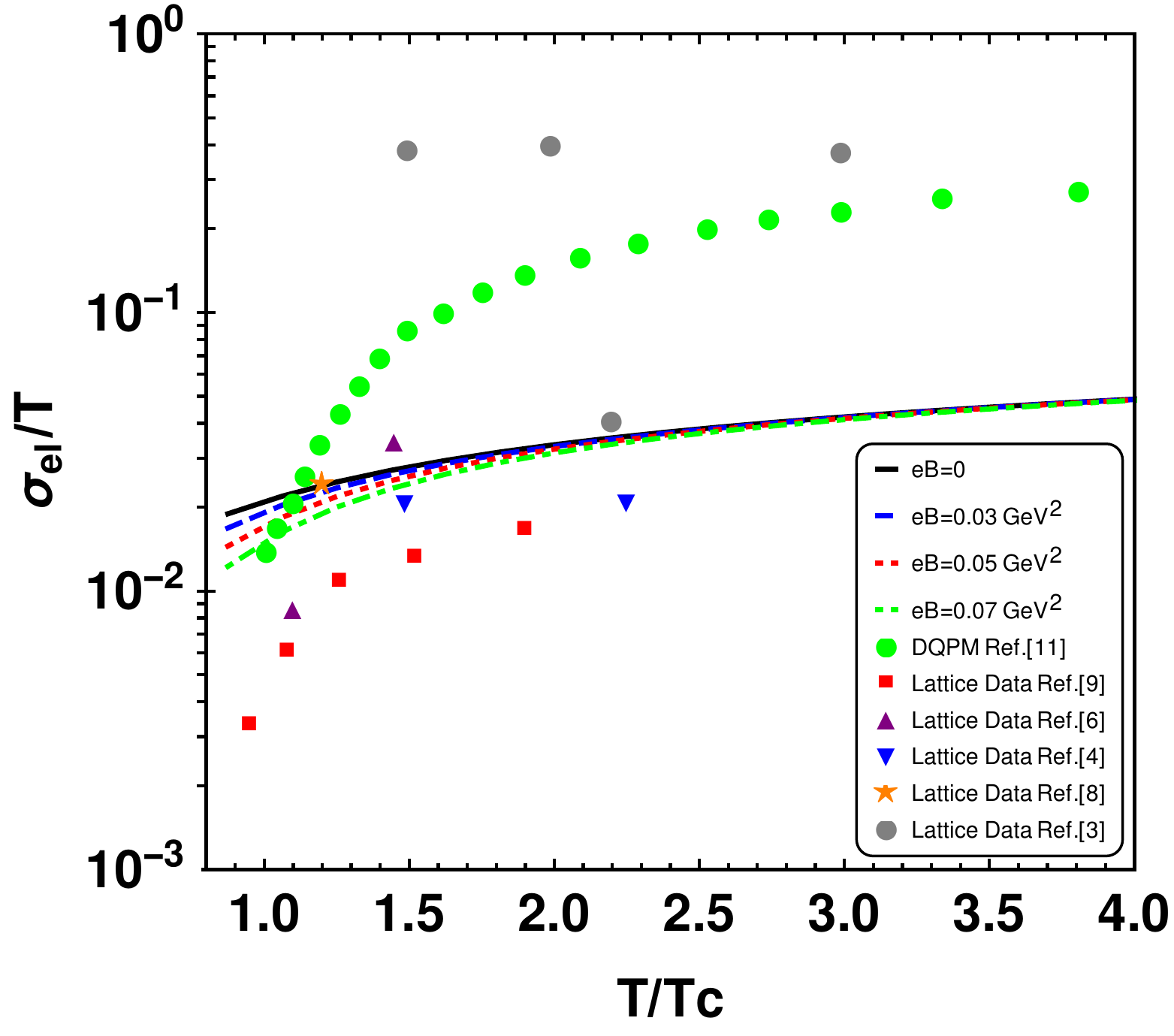}
\caption{Variation of $\sigma_{el}/T$ with respect to $T/T_c$ for different values of magnetic field 
	in the present calculation. Comparison with a different lattice result is also shown.}
\label{fig1}
\end{figure}
In Fig.~\ref{fig1}, we have shown the variation of the ratio of electrical conductivity to temperature ($\sigma_{\rm{el}}/T$) [Eq.~(\ref{siginB})] with respect to $ T/T_c $ at zero chemical potential for different values of magnetic field (i.e., $ B=0,0.03,0.05,0.07 $ GeV$ ^{2} $ etc.). Here we take $ T_c=160 $ MeV as the critical temperature corresponding to the quark-hadron phase transition. We found that the electrical conductivity decreases in the presence of the magnetic field. This shows that the system is electrically less conductive in the presence of magnetic field, particularly at low temperatures. Quarks experience a Lorentz force in the presence of magnetic field, which changes the moving direction of the particles. Thus, the electric current (flow of electric charge) carried by quarks in the plasma decreases in the direction of electric field. Hence the electrical conductivity which is proportional to the current in the direction of the electric field also decreases. 
We have compared our model results with the various lattice  calculations~\cite{Amato:2013naa,Ding:2010ga,Aarts:2007wj,Brandt:2012jc,Gupta:2004} and dynamical quasiparticle model (DQPM) results (green points)~\cite{Cassing:2013iz}.

\begin{figure}
	\includegraphics[scale=0.45]{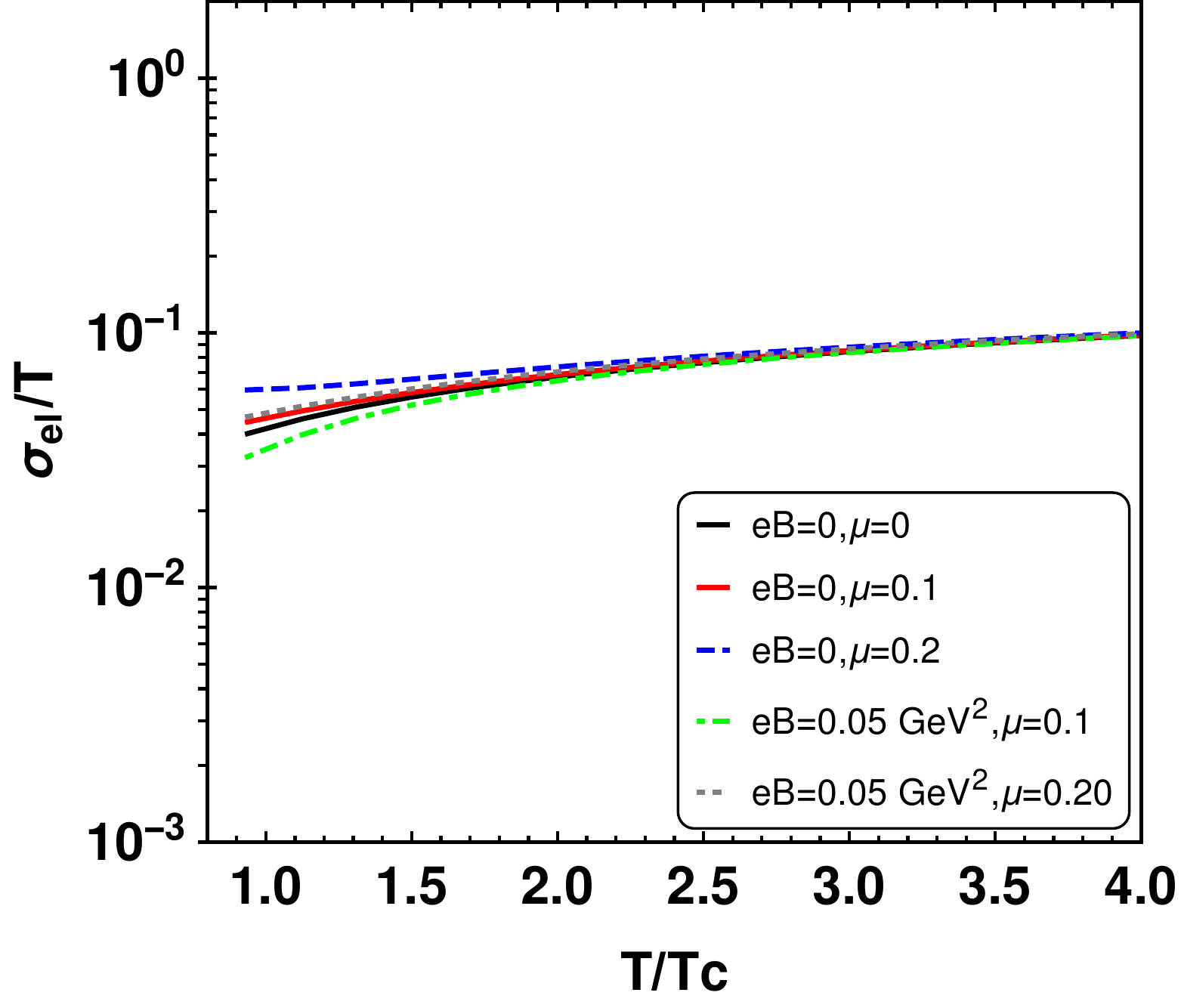}
	\caption{Variation of $\sigma_{el}/T$ with respect to $T/T_c$ for different values of magnetic field and chemical potential 
		in the present calculation.}
	\label{fig2}
\end{figure}

 Figure~\ref{fig2} shows the variation $\sigma_{el}/T$ with respect to temperature at finite chemical potential i.e.  $ \mu = 0, 100$ and $200$ MeV for both in the presence and absence of magnetic field.
 From the figure we observe that electrical conductivity is large at lower temperatures as compared to higher temperatures even in the presence of magnetic field. This effect of finite chemical potential is due to a sizable change in the distribution function of quarks at lower temperatures since the ratio $\mu/T$ is significant. On the other hand at higher temperature the ratio ($\mu/T$) becomes small as the temperature increases. Also the effect of magnetic field is much less. Therefore, the effect of finite chemical potential is much less on the distribution function and hence on the electrical conductivity as well even in the presence of magnetic field.

\section{Conclusion}
In this work, we have studied the electrical conductivity of the static QGP medium in the presence of constant and homogeneous external electromagnetic field. The relativistic Boltzmann's kinetic equation has been solved in RTA to calculate the electrical conductivity for the static QGP phase. Further, we have discussed the quasiparticle model and one-loop strong coupling constant in the presence of magnetic field. 
We have shown the variation of $\sigma_{\rm{el}}/T$ with respect to $ T/T_c $ 
in the presence of $ B $. We found that the electric conductivity decreases with the increase in magnetic field, especially at a low temperature. We have compared our results with the lattice as well as the dynamical quasiparticle model. We have observed that $ \sigma_{el} $ at a nonzero $ B $ remains within the range of the lattice and model estimate at $ B\neq 0 $. Finally we have extended our calculation to the finite chemical potential. We found that electric conductivity increases with increase in chemical potential at low temperature even in the presence of magnetic field.

{\bf Note added:} Recently, another paper by Das {\it et al.}~\cite{Das:2019ppb}, in which the electrical conductivity is studied in the presence of magnetic field using a quasiparticle model, appeared. However, we have use the one-loop strong coupling constant in the presence of magnetic field, whereas the authors use the two loop coupling constant without magnetic field.

\noindent
\section{Acknowledgments}
L.T. would like to thank Najmul Haque for constant support and help during the course of this work. L.T. was supported by National Institute of Science Education and Research (NISER), India, under institute postdoctoral research grant. PKS was supported by IIT Ropar, India, under institute postdoctoral research grant.


\end{document}